\documentclass[12pt,preprint]{aastex}
\newcommand{\B}[1]{\mbox{\boldmath $#1$}}
\newcommand{\BB}{\B{B}}
\newcommand{\Bx}{\B{x}}

\newcommand{\Bk}{\B{k}}

\newcommand{\Br}{\B{r}}
\newcommand{\BV}{\B{V}}

\begin{document}

\shorttitle{Anisotropy in the solar wind}
\shortauthors{Bigazzi et al.}
\title{Small-scale anisotropy and intermittency in high and low-latitude solar wind}

\author{A. Bigazzi\altaffilmark{1}}
\affil{Departamento de Matem\'atica Aplicada, Universidade do Porto,
Portugal}

\author{ L. Biferale }
\affil{
Dipartimento di Fisica and INFN, Universit\`a di Roma Tor Vergata, Italy. 
}
\author{S.M.A.Gama}
\affil{CMUP and Departamento de Matem\'atica Aplicada, Universidade do Porto,
Portugal}
\and
\author{M.Velli\altaffilmark{2} }
\affil{Jet Propulsion Laboratory, California Institute of Technology,
     CA USA}
\altaffiltext{1}{
Also visiting researcher, Dipartimento di Fisica, Universit\`a di Roma Tor
Vergata, Italy. 
}
\altaffiltext{2}{
{Also Dipartimento di Astronomia e Scienza dello Spazio,
Universit\`a di Firenze, Italy}
}
\begin{abstract}
We analyze low and high--latitude fast solar wind data from the Ulysses
spacecraft from 1992 to 1994 using a  
 a systematic method to analyse the anisotropic content of
the magnetic field fluctuations. 
We investigate  all  available frequencies,  $1-10^{-6}{\rm
Hz}$, for both high and low--latitudes datasets and are able to   
 quantify    the relative
importance of the anisotropic versus the isotropic  
fluctuations.   We analyse, up to sixth order,    longitudinal, transverse and mixed  
magnetic field correlations.
Our results  show  that strongly 
intermittent and anisotropic events  are present in  
  the solar wind plasma at high frequencies/small scales, 
indicating the absence of a complete recovery  of isotropy. 
Anisotropic scaling properties
are compatible for  high and low--latitude data,
suggesting a universal behaviour in spite of the different rate of 
evolution of the
fast solar wind streams in the two  environments. 
\end{abstract}
\section{Introduction}
The solar wind is an  inhomogeneous, anisotropic and  compressible 
 magnetized plasma where both velocity
and magnetic fields fluctuate over a broad range of frequencies and
scales,  see e.g. the reviews of \citet{TM95-R} and \citet{HT-R}.
Fluctuations may originate either from  the nonlinear 
interactions between large-scale streams
\citep{Col66,Col68,Matth+90}
or by interacting  Alfv\'en waves produced
close to the Sun and carried by the wind, 
\citep{BD71,DMV80,Leam+98}.
%
Observations of  the radial evolution  
of magnetic fields in the inner heliosphere show 
the presence of fully developped turbulent spectra
within a range of 
$10^{-4}-10^{-1}$~Hz  \citep{Bav82}.
%

%
The spectral index depends on the frequency range and on the distance
from the Sun, varying from  $-1.2$ to $-1.7$.
Low-frequency measurements are performed at around
$10^{-5}-10^{-2}$~Hz \citep{Col68}, 
while high frequency  measurements 
sample the range closer to $10^{-2}-10^{-1}$~Hz, 
\citep{Bav82,Leam+98,Horbury+01}.
The spectral index tends to flatten closer to the Sun, indicating that
turbulence is evolving in the solar wind. 
Phenomenological theory of 
hydrodynamic turbulence  \citep{K41}
predicts a value of  $-5/3$ for the
spectral index, while the theory  of  
Alfv\`en wave-driven magnetohydrodynamic (MHD) turbulence of 
\citet{I63}  and \citet{Kr65},
predicts a slope of  $-3/2$. 
Neither prediction takes into account 
the possbile influence of anisotropies and the presence of intermittency
\citep{Burlaga91,Burlaga92,MarschLiu93,Carb93,FeyR94,CarbVB95,Horbury+97,
Ruzmaikin95,BCSB03,Hnat+03,BershSr04},
in a systematic way. 

The presence of anisotropy  makes it difficult to compare observed data 
with the two predictions, while the presence of intermittency 
tells us that the characteristics of the spectrum are not sufficient to
characterize the system:
   higher order statistics need to be taken into
account. 
In particular, spectral indices alone are insufficient to discriminate amongst
turbulence models.  

Anisotropy has been measured by various techniques involving the calculation of 
 second order 
moments of the field either in the real or Fourier space, such as 
the variance matrix or the power spectra 
\citep{BD71,CMV95}.
The  eigenvector of the variance matrix corresponding  to
the minimum eigenvalue is usually known as the minimum variance direction.
This direction is aligned with the large--scale mean field,  indicating
suppression of turbulence in that direction  
\citep{Leam+98,Bruno+99}.
Several MHD models incorporate at various levels  the  asymmetry of the
spectral indices in the field-aligned (longitudinal) and transverse directions
\citep{SMM83,ZM92,NB96,GS97,MOGH98}.
Although this is a possible way to characterize anisotropy, 
second order longitudinal and transverse structure functions 
contain both anisotropic and isotropic contributions, as will be detailed
later in Section~\ref{s:aniso}. 
Those two contributions, always mixed, need proper treatment to
be disentangled. 
A more systematic approach  to analyse anisotropy is therefore important. 
Moreover, the relation between anisotropy and intermittency has not been 
investigated so far. 
Anisotropy and intermittency may also be important in the context of  
scattering of particles in the heliosphere (see e.g. \citet{GJ96}) 

We present in this paper a  method for  extracting in a systematic way,
from the one-dimensional
spacecraft data,  information  on the anisotropy 
and intermittency  of the  magnetic field fluctuations, 
and the interplay between them. 
We base our  analysis on the behaviour of both 
diagonal and non-diagonal components  of  higher order 
structure functions. 
We have systematically compared isotropic and
anisotropic fluctuations at different scales and for different
magnetic correlation functions. 
We measure how fast isotropy is recovered at small scales, concerning
both typical fluctuations of the order of the mean standard deviation,
and highly intermittent events, affecting more the tails of the
magnetic field probability density at all scales.
We use Ulysses data of high speed streams at two different 
points 
along its orbit, at high and low latitudes, in order to
assess the dependence on the large-scale properties of the
small-scales anisotropic fluctuations, i.e. the issue of small-scales
universality.  

 The paper is organized as follows. In Section~\ref{s:aniso} we
present the set of observables needed to have a systematic
control on the isotropic and anisotropic ensembles.  In Section~\ref{s:data}
 we present
our data set and  in Section~\ref{s:res}
the main results for both the low and high latitudes data.
Section~\ref{s:concl},  summarizes our findings suggesting  
further  possible investigations. 
\section{Anisotropy and Structure function analysis} \label{s:aniso}
Structure function decomposition into 
isotropic and anisotropic components 
has already been exploited with success in hydrodynamics,
both for experimental and numerical data analysis
\citep{arad98,arad99,kurien00,biferale01,biferale01a,shen02},  
see also \citet{biferale04} for a recent review.
In the latter case,
the anisotropic contents of fully developed flow has been systematically
analyzed. Anisotropic fluctuations of the velocity field
 have been shown to be characterized by anomalous scaling, 
thus   explaining  the 
 higher than predicted anisotropy found in the
gradient statistics,  
an effect known as ``smal-scales persistence of anisotropies"
\citep{shen:00,shen:02,biferale01a}. 
Some analytical results on the persistency of anisotropies for magnetic fields
have also been obtained in the simplified case of passive magnetic advection
by stochastic velocity fields \citep{ref:01FGV,lanotte,ref:00ABP}.
The ideal way to assess the relative isotropic/anisotropic contents at
all scales is to perform a decomposition of the correlation
functions, of order 2 and higher, over a suitable eigenbasis with
definite properties under the group of three-dimensional 
spatial rotations (the SO(3) group), corresponding to  spherical harmonics 
decomposition for the simpler case of
scalars.  In principle, one needs to distinguish among different anisotropic
contributions corresponding to the different projections on the whole
set of eigenfunctions \citep{arad99}.
Spacecraft data are inherently one-dimensional,
therefore not directly suitable to be fed into an SO(3) analysis,  which
requires the whole field in a 3D volume to be systematically worked out.
However, 
 we shall show how it is possible to construct
correlation functions of different orders that have null projection
over the isotropic ensemble, i.e. with a leading contribution coming
only by its leading anisotropic content, if any
\citep{kurien00,wander,Jacob}.  

Data analysis is based on a set of multi-scale
correlation functions built  upon different combinations of magnetic
field components.  The most general $n$th order correlation, 
 $S^{(n)}_{\alpha_1,\dots, \alpha_n}(\Br)$, 
 depending on
single separation $(\Br)$,
is built from  the $n$ spatial
increments of  magnetic field components:
\begin{equation}
S^{(n)}_{\alpha_1,\dots,\alpha_n}(\Br)=\langle 
\delta_{\Br}B_{\alpha_1}\delta_{\Br}B_{\alpha_2} \cdots
\delta_{\Br}B_{\alpha_n}
\rangle
\label{eq:s}
\end{equation}
where 
\begin{equation}
\delta_{\Br}B_{\alpha} \equiv B_{\alpha}(\Bx+\Br)-B_{\alpha}(\Bx)
\label{eq:del}
\end{equation}
is the difference between the values of component $B_{\alpha}$ at two
different points a distance $\Br$ away.  Brackets
$\langle\cdot\rangle$ in (\ref{eq:s}) indicate the average over the
locations $\Bx$. Notice that in (\ref{eq:s}) we have assumed
homogeneity but not isotropy, i.e. the correlation functions keep their 
explicit dependence on the full vector $\Br$.
The  correlation function (\ref{eq:s})  includes both  {\it isotropic}
 and {\it anisotropic} contributions:
\begin{equation}
S_{\alpha_1,\dots,\alpha_n}^{(n)}(\Br)=
S_{\alpha_1,\dots,\alpha_n}^{(n),iso}(\Br)+
S_{\alpha_1,\dots,\alpha_n}^{(n),aniso}(\Br).
\label{eq:s_aniso}
\end{equation}
In principle, different anisotropic contributions exist and one would need 
to further distinguish among them. In this study we limit ourselves to 
disentangling the isotropic contributions from the anisotropic, 
without entering
the more subtle problem of separating out all the different anisotropies
(the
interest reader may consult  \cite{biferale04}
 for a detailed illustration on how
to proceed in this direction). 

For $n=2$ and
$\alpha_1=\alpha_2$ in (\ref{eq:s}), we get the well known positively defined
second order structure function, connected to the the energy spectrum
$E_{\alpha,\alpha}(\Bk) = \langle |\hat{B}_{\alpha}(\Bk)|^2 \rangle $
via a Fourier transform.  Another widely used form of (\ref{eq:s}) is
the longitudinal structure function, obtained by projecting
all field increments along the separation versor, $\hat{\Br}$:
$S_L^{n}(r) = \langle
\left(\delta_{\Br}\BB \cdot \hat{\Br} \right)^{n} \rangle$. 
The  general form of the tensor
(\ref{eq:s}) for $n=2$ in the case of a fully isotropic and parity invariant
statistics,   is  given by the  combination
of the separation vector $\Br$ and the only isotropic second order
tensor, the unity matrix $\delta_{\alpha,\beta}$: 
\begin{equation}
 S_{\alpha_1,\alpha_2}^{(2),iso}(\Br)=\langle 
\delta_{\Br}B_{\alpha_1}
\delta_{\Br}B_{\alpha_2}
\rangle^{iso} = a(r) \delta_{\alpha_1,\alpha_2} + b(r) r_{\alpha_1} r_{\alpha_2}
\label{eq:s2}
\end{equation}
where $a(r)$ and $b(r)$ are two scalar functions depending only on
the amplitude $r = |\Br|$.  Similarly,  the expression for the fourth order isotropic tensors,
$S_{\alpha_1,\cdots,\alpha_4}^{(4), iso}(\Br)$,  comprises three scalar 
functions, $c(r),d(r),f(r)$:
\begin{eqnarray}
 S_{\alpha_1,\alpha_2,\alpha_3,\alpha_4}^{(4),iso}(\Br)
&=&\langle 
\delta_{\Br}B_{\alpha_1}
\delta_{\Br}B_{\alpha_2}
\delta_{\Br}B_{\alpha_3}
\delta_{\Br}B_{\alpha_4}
\rangle^{iso} = f(r) r_{\alpha_1}r_{\alpha_2}r_{\alpha_3}r_{\alpha_4} +\nonumber\\
&&c(r)  (\delta_{\alpha_1,\alpha_2}\delta_{\alpha_3,\alpha_4} + perm.) + d(r) 
(\delta_{\alpha_1,\alpha_2}r_{\alpha_3}r_{\alpha_4} + perm.)
\label{eq:s4}
\end{eqnarray}
Analogous expressions hold for higher oder isotropic correlation functions.
The key observation is
that by a suitable choice of the combination of indices
$\alpha_1,\cdots,\alpha_n$ and of the orientation $\Br$ one may have
the isotropic components vanish at any order, $n$ in
(\ref{eq:s_aniso}). 
From now on, let us fix the
separation distance in the direction $\hat{\Bx}$ so that $\Br = (r_x,0,0)$.
For the case $n=2$, when
$\alpha_1 \ne \alpha_2$, the resulting isotropic
components vanish.  We  therefore have three different second
order correlation functions that  are {\it purely anisotropic}.  
 When the order $n$ of the correlation
function is {\it even} it is enough to take an {\it odd} number of
field increments in two different directions to have a {\it purely
anisotropic} observable. 
Therefore a  possible set of purely anisotropic correlations have the form:
\begin{equation}
S_{\alpha,\beta}^{p,q}(r_x)=\langle 
\delta_{r_x}B_{\alpha}^p
\delta_{r_x}B_{\beta}^{q}
\rangle \quad (p+q=n)
\label{eq:s1}
\end{equation}
with both $p$ and $q$ odd and such that $p+q=n$.
  The above $n$th order correlation has a vanishing isotropic
component when the combinations of indices $\alpha = x$ and 
 $\beta= y,z$ are taken. 

Before presenting the results of our data analysis, let us briefly
comment on the translation from time series to spatial signals in our
dataset. Of course, as it is the case for all spacecraft data, we only
have access to the time evolution of the magnetic field along the
orbit. We therefore cannot make an explicit evaluation of simultaneous
field increments over space. Nevertheless, the advecting velocity
speed is so high (see next section for a summary of the main physical
relevant quantities) that in the range of frequencies we are
interested in, it is possible to  safely adopt the Taylor hypothesis and
 translate time
increments into spatial increments. 
The ``Taylor hypothesis'' consists in supposing  the 3D
field as frozenly advected by the underlying large scale velocity
field, $\BV_0$ \cite{Frisch95}.
 Field increments in  the
same spatial point at two times, $t,t'$, are considered  equal to the
instantaneous field increments over two spatial locations $\Bx$ and
$\Bx+\Br$ with $\Br = \BV_0 \, (t'-t) $.  Therefore, for us, the direction
$\Br$ is fixed and given by the direction of the wind at the location of the
spacecraft that is, within a few percent, the spacecraft-Sun
direction. This direction, as said before, will be taken as our
reference $\hat{\Bx}$ axis. Spatial homogenity is translated via the Taylor hypothesis into temporal stationarity.

\section{Ulysses Dataset}
\label{s:data}
\begin{deluxetable}{cccccc}
\tablewidth{0pt}
\tablecaption{
Low latitudes and Polar datasets.
\label{tab1}
}
\tablehead{
\colhead{Dataset}&
\colhead{Days}&
 \colhead{Lat  } & 
\colhead{Dist  }&
\colhead{Speed }&
\colhead{$\langle B\rangle$  } \\
\colhead{}&
\colhead{}&
 \colhead{ (HGL)}& 
\colhead{ (A.U.)}&
\colhead{(Km\, s$^{-1}$)}&
\colhead{(nT)}
}
\startdata
Low lat& 
92/209-93/137&
-15 to -30&
5.3 to 4.7 &
750&
0.47 \\
Polar & 
94/245-265&
-79.7 to -80.2&
2.37 to 2.23&
760&
1.3\\
\enddata
\end{deluxetable}
%
Ulysses orbit samples the interplanetary plasma at distances varying
approximately from 1 to 6 A.U, on a polar orbit. It is therefore
possible to follow the evolution of plasma characteristics with
distance and latitude.  We use two different set of data: the
first one was taken by Ulysses during 1992-1993, when the spacecraft
was at about $20^{\circ}$ heliographic latitude and 5 AU distance from
the Sun. The second was taken at the end of 1994, 
with Ulysses above the South Pole, at about $80^{\circ}$ latitude and 
a distance
of about 2 AU from the Sun. 
This latest dataset has just
recently been made available to the community by the Ulysses team.
Solar activity was, during the 92-93 period, declining, after the 1990
maximum. In 1994, the cycle was approaching the minimum of 1996.  Each
daily dataset provides the magnitude of all three components of the
interplanetary magnetic field, taken at the rate of 1 or two seconds by the
Vector Helium Magnetometer on board (Balogh et al., 1992).
In  Table~\ref{tab1} we report, for the two datasets, the interval of time
considered, the heliographic latitudes spanned, distance from the Sun, 
average speed of the wind and average magnetic field intensity. 

We pre-process data  in order to clean spikes due to
instrumental problems or to large shocks. This is made by excluding
those data where the jump in the magnetic field between two
consecutive data points (usually 1 second apart)
is larger than a threshold, $\Delta B$, 
of the order of the mean large scale
magnetic field. A fraction of datapoints as small as 
$10^{-5}$, is discarded this way. 
 As a result,  we
can access  magnetic field fluctuation on a range of frequencies of
almost six  decades. \\
In  Table~\ref{Tab:thresh} we detail the 
total number of datapoints in the dataset, the  number of datapoints
discarded,
$N_{\rm excl}$, the fraction of the latter to the total, the 
 threshold on the maximum jump between magnetic field for 
consecutive datapoints, the average field intensity for the whole dataset.
\begin{deluxetable}{cccccc}
\tablewidth{0pt}
\tablecaption{Data selection 
\label{Tab:thresh}
}
\tablehead{
\colhead{Dataset } &
\colhead{$N$} &
\colhead{$N_{\rm excl}$} &
\colhead{$N_{\rm excl}/N$} &
\colhead{$\Delta B$} &
\colhead{$\langle B\rangle$}
\\
\colhead{} &
\colhead{} &
\colhead{}&
\colhead{}&
\colhead{$(nT)$}&
\colhead{$(nT)$}
}
\startdata
Low latitude & 3 915 792& 78&
2.0\ E-5& 0.5& 0.47
\\
\hline 
High latitude & 1 476 051& 36&
2.4\,E-5& 1.2& 1.30
\enddata
\end{deluxetable}
%
\subsection{Low latitude  dataset}
\begin{figure}
\plotone{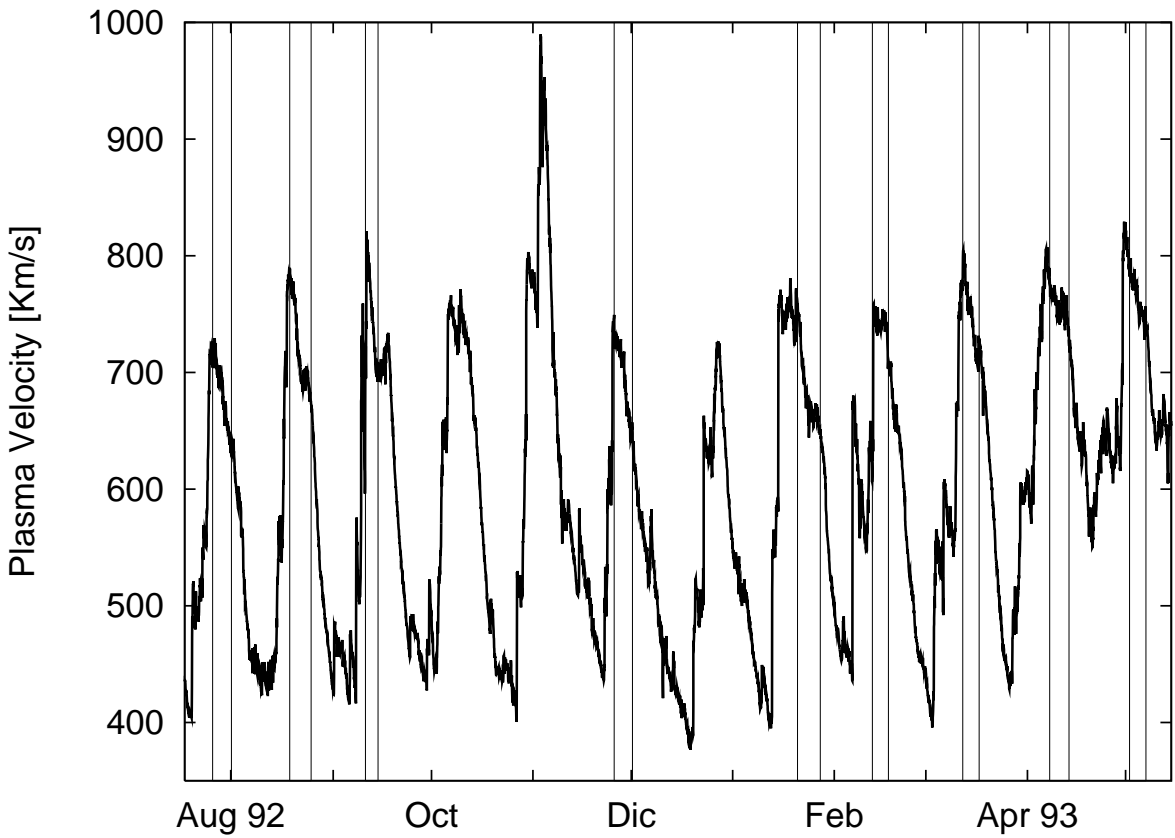}
\caption{ 
Plasma velocity sampled by Ulysses spacecraft between day 209 (July
27) 1992 and day 137 (May 17) 1993. Spacecraft  was between
$-15^{\circ}$ 
to $-30^{\circ}$ 
heliographic latitude, approaching the Sun
 at a distance varying from 5.3 to 4.7 AU (see Table~\ref{tab1}).
Vertical lines highlight selected intervals in the trailing edges of
high--speed streams. 
}
\label{fig1}
\end{figure}

The alternating pattern of slow and fast wind is shown in 
Fig.~\ref{fig1}, spanning a ten month period, from day 209,
1992, to day 137, 1993. Within this period, we selected those sequences,
of about five days each, when spacecraft is embedded in the trailing edges of 
high speed streams and velocity is above 650 Km/s. 
The days selected are, in 1992, 209-214, 235-241, 259-263, 337-342 and,  
in 1993,  28-34, 53-57, 81-85, 108-113, 133-137. They are highlited in
Fig.~\ref{fig1} within vertical lines.  
\subsection{High latitude dataset}
Twenty-one consecutive days around the maximum latitude reached at perihelium,
during the fast latitude scan of 1994, are selected. Differently
from the previous dataset, only
the fast component of the wind is present.
Table~\ref{tab1} lists latitude range, distance, average speed and average
magnetic field for this dataset as well. 
\section{Results}
\label{s:res}
\subsection{Equatorial data}
\label{s:equat}
\begin{figure}
\plotone{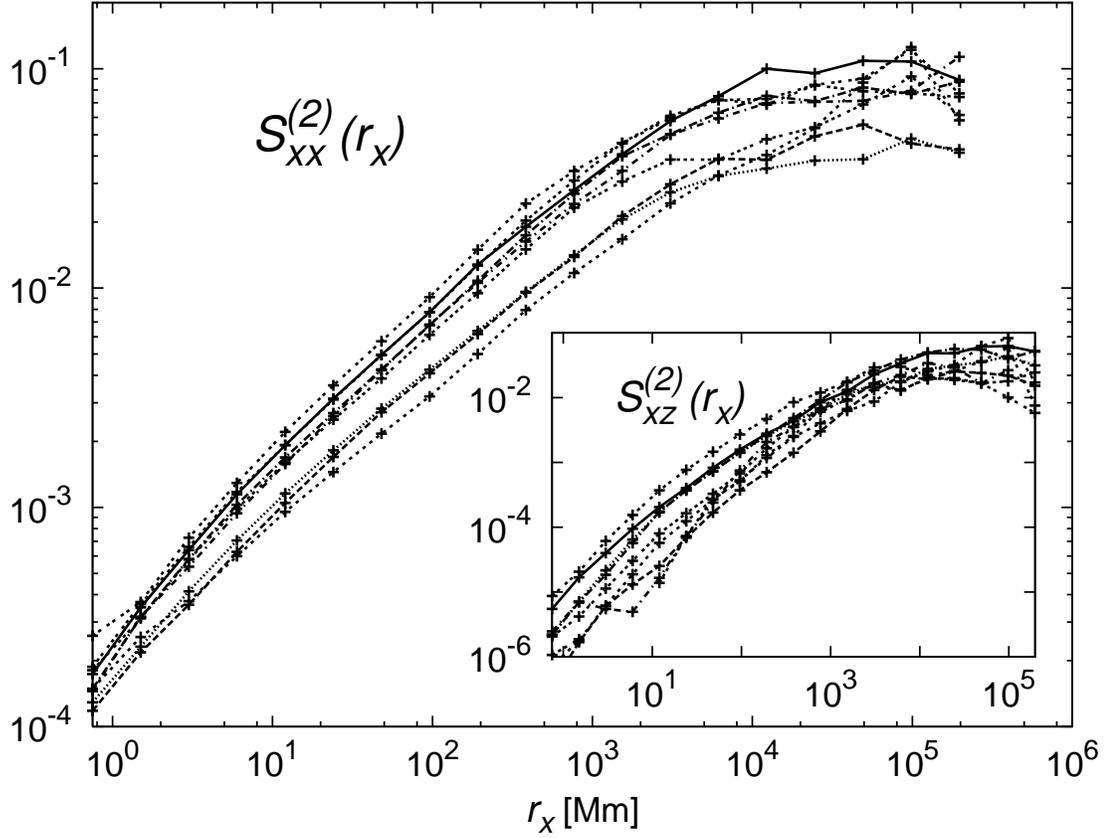}
\caption{
Second order longitudinal structure function, $S^{(2)}_{xx}(r_x)$, for 
each interval comprising the low latitude dataset (see Fig.~\ref{fig1}),   
as a function of the separation $r_x$.
In the inset, the second order purely anisotropic structure function, 
$S^{(2)}_{xz}(r_x)$. 
}
\label{fig2} 
\end{figure}

We want to first test the consistency between the disjoint sets 
 making up the low latitude dataset of  Fig.~\ref{fig1}. 
The second order longitudinal structure
functions, calculated for each  of those intervals of contiguous data, are
shown in  Fig.~\ref{fig2}. 
 They  are consistent with each other over more than 5 decades, from 
 $1$ to $10^5$~Hz in the spacecraft frame,  which translates,
with a mean plasma velocity of $750$~Km/s, into  a range of 
 $7.5 \cdot 10^{-1}$~Mm to $7.5 \cdot 10^{4}$~Mm.  
Some intervals have  a more intense signal than others do. 

The anisotropic component $S^{(2)}_{xz}$ shown in the inset of the same figure, 
 displays a similar
behavior. We conclude that data from different intervals
are commensurable and  combine them together to obtain
more stable statistical results.  We shall  refer to the combined set 
as the ``low-latitude" dataset without further distinction. 

%
%

\begin{figure}
\plotone{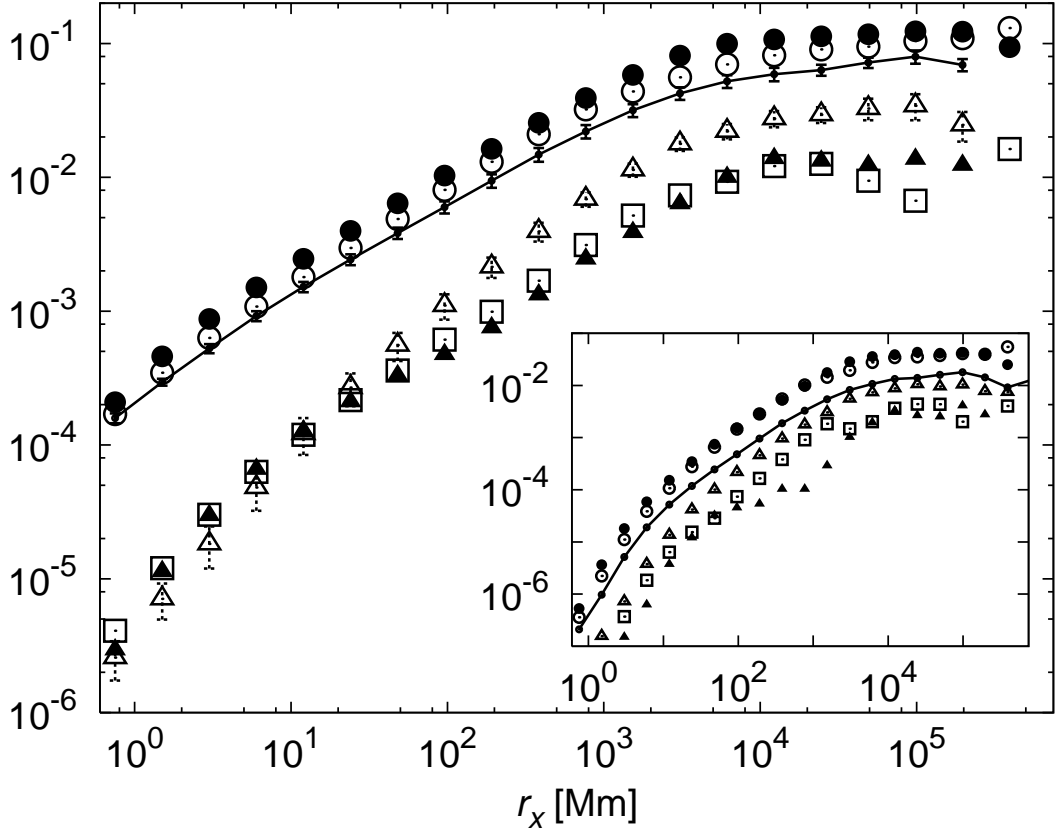}
\caption{Second order longitudinal, transverse and {\it purely 
anisotropic}
structure functions. Low latitude dataset.
The upper three curves show the longitudinal and tranverse structure
functions: 
solid line ---  $S^{(2)}_{xx}$;  
empty cirles $\circ$  $S^{(2)}_{yy}$; 
filled circles $\bullet$ $S^{(2)}_{zz}$. 
Errorbars are superimposed on --- $S^{(2)}_{xx}$. Errors are evaluated as the
standard deviation  of the individual intervals comprising the whole dataset. 
The lower curves show the purely anisotropic structure functions:
$S^{(2)}_{xy}$, $\blacktriangle$ filled triangles; 
$S^{(2)}_{xz}$, $\vartriangle$ empty triangles; 
$S^{(2)}_{yz}$, $\square$ empty squares. 
Errorbars are superimposed on  $\vartriangle$ $S^{(2)}_{xz}$. 
Inset: fourth order structure functions, longitudinal, transverse and {\it
purely anisotropic}. 
Solid line, --- $S^{(4)}_{xxxx}$; 
empty circles $\circ$ $S^{(4)}_{yyyy}$; 
filled circles $\bullet$ $S^{(4)}_{zzzz}$. 
{\it Purely anisotropic} structure functions are:
$S^{(4)}_{xyyy}$, $\blacktriangle$ filled triangles; 
$S^{(4)}_{xzzz}$, $\vartriangle$ empty triangles;
$S^{(4)}_{yzzz}$, $\square$ empty squares.
}
\label{fig3}
\end{figure}

Let us now compare the  undecomposed  second order structure
functions with its anisotropic content.  
In Fig.\ref{fig3} we plot the
longitudinal structure functions of second order, $S^{(2)}_{x,x}(r_x)$ and
the two  transverse structure functions in the directions perpendicular
to the $\hat{\Bx}$ axis, 
$S^{(2)}_{yy}(r_x)$ and $S^{(2)}_{zz}(r_x)$.
All  these functions have both isotropic and anisotropic contriution:
\begin{equation}
S_{\alpha,\alpha}^{(2)}(r_x) = S_{\alpha,\alpha}^{(2),
iso}(r_x)+S_{\alpha,\alpha}^{(2),aniso}(r_x)
\label{eq:sl}
\end{equation}  
The two {\it purely anisotropic} second order structure
functions $S_{xy}^{(2)}(r_x)$ and $S_{xz}^{(2)}(r_x)$, are plotted in the same
figure. 
A few comments
are in order. First, we notice that the {\it anisotropic} correlations
have a smaller amplitude with respect to the full correlation
functions. This suggests that the isotropic contribution in the
decomposition (\ref{eq:s_aniso}) is  dominant. Moreover, we see
that the anisotropic curves decay slightly faster than the full correlation by
decreasing the  scale. In other words, isotropic
fluctuations become more leading going to small
scales, but very slowly. This is consistent with  the {\it
recovery-of-isotropy} assumption often advocated in many
phenomenological theory of hydrodynamic turbulence and magnetized
plasma. However, in order to assess more precisely this issue, it is important 
to
control higher order statistical objects, i.e. the whole shape of
the probability density distribution, at all scales. In the inset of
the same Fig.~\ref{fig3} we show the same comparison between
longitudinal, $S_{xxxx}^{(4)}(r_x)$, transverse,
$S_{\alpha\alpha\alpha\alpha}^{(4)}(r_x)$ (with $\alpha=y,z$) and {\it purely
anisotropic} correlations of {\it fourth order} (see caption in the
figure).  Now the situation is quite different. First, the intensity of
some
{\it purely anisotropic} components are much closer to those with mixed 
 isotropic and anisotropic contributions, i.e. the longitudinal and
transverse structure functions.  Second, the decay
rate as a function of the scale is almost the same: no recovery of
isotropy is any more detected for fluctuations of this order. A
similar, even more pronounced, trend is observed for sixth order
quantities (not shown). The persistence of strong anisotropies at
high frequencies (small scales)  cast some caveat on measurements
of quantities which do not properly disentangle the  isotropic  from the 
anisotropic components. 
As it will be shown later for the case of high-latitudes data, 
anisotropic components have 
strong variations in intensity depending on
the position on the solar orbit. Both latitude and distance from the
Sun  influence the amount of anisotropy.
As a result, undecomposed quantities which are influenced by both isotropic and anisotropic
fluctuations are expected to be non-universal, the anisotropic content being
dependent on the spacecraft position and latitude.  
This must hold for the  spectrum  and even more for 
higher order structure functions. 
%
\subsection{Intermittency}
\begin{figure}
\plotone{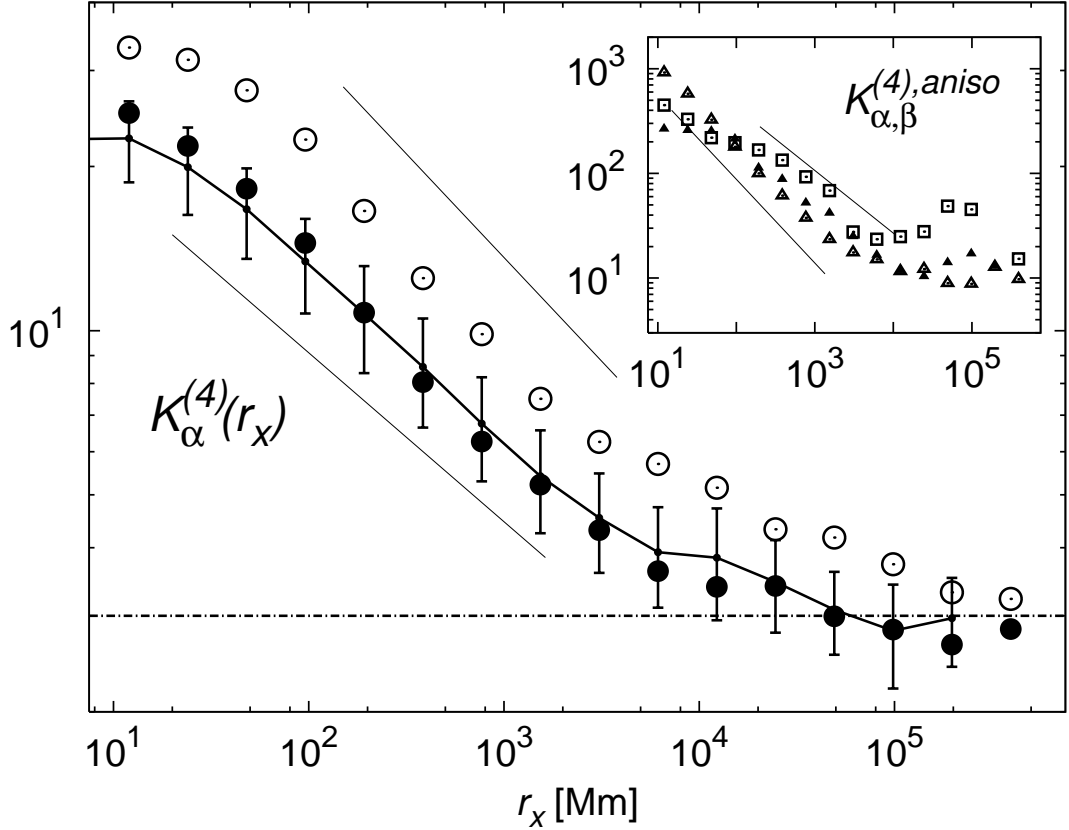}
\caption{
Kurtosis (\ref{eq:kurt_ls}) of longituinal and transverse magnetic field
fluctuations.
Solid line --- $K^{(4)}_{x}(r_x)$; 
empty circles $\circ$ $K^{(4)}_{y}(r_x)$;
filled circles $\bullet$ $K^{(4)}_{z}(r_x)$.
Straight lines represent linear fit to the central portion of the 
curves with slopes  of $ -0.31$ for the longitudinal component, and
$ -0.38$ for the two transverse ones. 
The horizontal line corresponds to the Gaussian value of $3$, attained only at
large scales. 
In the inset: {\it purely anisotropic} kurtosis (\ref{eq:kurtanis}). 
$K^{(4)}_{xy}(r_x)$,   $\blacktriangle$ filled triangles;
$K^{(4)}_{xz}(r_x)$,  $\vartriangle$ empty triangles;
$K^{(4)}_{yz}(r_x)$,  $\square$ empty squares.
Straight lines have   
slopes of  $\chi_4^{aniso}=-0.8 $ and  $\chi_4^{aniso}=-0.6 $ in the 
$xz$ and $xy$ components, respectively.   Low latitude dataset. 
}
\label{fig4}
\end{figure}

Anisotropic fluctuations are not the unique source of
complexity in solar wind data. It is well known that both magnetic and
velocity fields are strongly intermittent, i.e. their statistical
properties at different scales cannot be simply superimposed by 
rescaling. This implies the existence of
anomalous scaling laws in the structure functions and ``fat tails'' in
the PDFs of field increments \cite{Frisch95}. Here we want to address this
issue  for  the anisotropic sectors.
The main conclusion will be that 
anisotropic correlations also show anomalous scaling, their
PDFs becoming more and more non-Gaussian at small 
scales. 
 In Fig.~\ref{fig4}
we show the Kurtosis of both the longitudinal and transverse structure
functions, i.e. the ratio between fourth order moments and square of
the second order moments of longitudinal and transverse increments:
\begin{equation}
K^{(4)}_{\alpha}(r_x) =
\frac{S_{\alpha\alpha\alpha\alpha}^{(4)}(r_x)}{
(S_{\alpha\alpha}^{(2)}(r_x))^2} 
\label{eq:kurt_ls}
\end{equation}
A Gaussian variable would have a Kurtosis of $3$, independent on the scale
while all three curves  grow at small scales.  
We stress once more here  that these quantities probe both the 
 iso and anisotropic physics.  Therefore 
 the scaling properties are certainly affected by 
 the superposition of different contributions.
In the previous section we have shown that the 
isotropic sector is never sub-leading.  We may therefore consider the
above result as a confirmation that the isotropic fluctuations are 
indeed strongly intermittent. 

Similarly, to investigate intermittency in the anisotropic sector, 
it is useful to define a
{\it purely anisotropic} Kurtosis, by taking the adimensional ratios of 
fourth order and second order
 anisotropic correlation functions:
\begin{equation}
K_{\alpha\beta}^{(4),aniso}(r_x) =\frac{S_{\alpha\beta\beta\beta}^{(4)}(r_x)}
{(S_{\alpha\beta}^{(2)}(r_x))^2} \sim r_x^{\chi_4^{aniso}}
\label{eq:kurtanis}
\end{equation}
where $\alpha,\beta$ are chosen so that  contributions from 
the  isotropic sector in 
both the numerator and the denominator vanish.  
The anisotropic components of the kurtosis (\ref{eq:kurtanis})
are shown in the inset of Fig.~\ref{fig4}. 
Functions  are increasing 
 towards small scales, with slopes of  
$\chi_4^{aniso} = -0.6 \pm 0.2 $, 
$\chi_4^{aniso} = -0.8 \pm 0.2 $, 
$\chi_4^{aniso} = -0.45 \pm 0.2 $ 
for the $xy$, $xz$ and $yz$ components, respectively (see
Table~\ref{tab:univ}).
This is the first clear indication, to our knowledge, that
 anisotropic fluctuations in the solar plasma are strongly
 intermittent. 
Similar
 trends are observed for generalized Kurtosis of sixth order (not
 shown):
\begin{equation}
K_{\alpha\beta}^{(6),aniso}(r_x) =\frac{S_{\alpha\alpha\alpha\beta\beta\beta}^{(6)}(r_x)}
{(S_{\alpha\beta}^{(2)}(r_x))^3} \sim r_x^{\chi_6^{aniso}}
\label{eq:kurtanis6}
\end{equation}
 There, our best estimate for the exponents is
 $\chi_6^{(aniso)} = -1.2 \pm 0.3$, $xy$ component, and 
 $\chi_6^{(aniso)} = -1.5 \pm 0.3$, $xz$ component.

Let us here remark that the
 quantity in (\ref{eq:kurtanis}) is not constructed from  ratios of 4th
 and 2nd order moments of the same observable, i.e. it is not,
 rigorously speaking, the kurtosis of a stochastic
 variable. Nevertheless, it is a good probe of the relative intensity of
 4th versus 2nd order anisotropic moments, 
the best  that can be done with a one-dimensional set of data.

A power law fit of the numerator and denominator of
(\ref{eq:kurtanis},\ref{eq:kurtanis6}) 
can be used  to directly measure the scaling exponents of the second order,
\begin{equation}
S_{\alpha\beta}^{(2)}(r_x) \sim r_x^{\zeta_2^{(aniso)}}, 
\label{eq:zeta2}
\end{equation}
and higher order anisotropic correlation functions, 
\begin{equation}
S_{\alpha\beta\beta\beta}^{(4)}(r_x) \sim r_x^{\zeta_4^{(aniso)}}
\qquad
S_{\alpha\alpha\alpha\beta\beta\beta}^{(6)}(r_x) \sim r_x^{\zeta_6^{(aniso)}}
\label{eq:zeta4}
\end{equation}
with, as customary now,  $\alpha,\beta$ are chosen in such a way  that 
only {\it purely anisotropic} quantities are returned.  We found 
$\zeta_2^{(aniso)} = 0.75 \pm 0.1 $ for the $xy$ component,
$\zeta_2^{(aniso)} = 0.95 \pm 0.1 $ for the $xz$ component, and
$\zeta_2^{(aniso)} = 0.75 \pm 0.1 $ for the $yz$ component, see
Table~\ref{tab:univ}. 
Values for the fourth and sixth orders $\zeta_4^{(aniso)}$ and 
 $\zeta_6^{(aniso)}$  may also be read out from the same table. 
Errorbars are estimated from the change of slope in the range of scales from
$10$ to $10^3$~Mm. Missing entries in the table
 indicate that the scaling properties
were not well defined within that range. 

\begin{figure}
\plotone{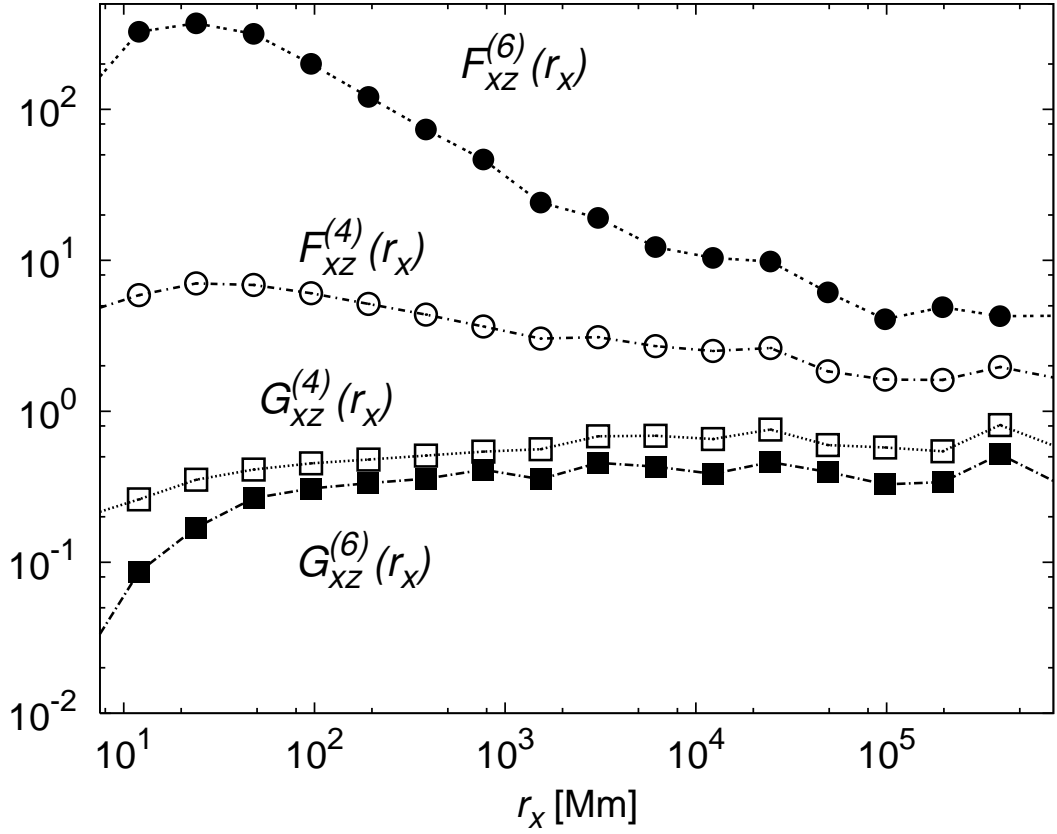}
\caption{
Generalized flatness $G^{(n)}_{\alpha\beta}(r_x)$ and
$F^{(n)}_{\alpha\beta}(r_x)$ of order 4 and 6, 
Eqs.~(\ref{eq:kurtnew}) and (\ref{eq:kurtmixed}) for components $xz$
and $xy$.
Low latitude dataset.}
\label{fig5}
\end{figure}

The above results show that anisotropic fluctuations, 
although they never become  the leading ones are  still important at small
scales. Order by
order, the undecomposed correlation function is more intense than
any anisotropic projection. This can be visualized, for
the $4$th and $6$th orders, by plotting the
ratio between the undecomposed object and one anisotropic projection:
\begin{equation}
G^{(4)}_{xz}(r_x) = \frac{S_{xzzz}^{(4)}(r_x)}{S_{xxxx}^{(4)}(r_x)};
\qquad G^{(6)}_{xz}(r_x) = \frac{S_{xxxzzz}^{(6)}(r_x)}{S_{xxxxxx}^{(6)}(r_x)}
\label{eq:kurtnew}
\end{equation}
These quantities never increase at small scales, indicating 
that isotropic contribution in the denominator is leading
with respect to the anisotropic,  see Fig.~\ref{fig5}.
Another quantity  that can be used to  characterize  the relative 
weight of anisotropic to isotropic fluctuations, may  be built  from a $n$th
order anisotropic moment and the $n/2$ power  of a 2nd order isotropic
moment \citep{shen02,biferale01a}. For example, in our geometry, one
possible choice would be:
\begin{equation}
F^{(4)}_{xz}(r_x) =
\frac{S_{xzzz}^{(4)}(r_x)}{(S_{xx}^{(2)}(r_x))^2}
\qquad  F^{(6)}_{xz}(r_x) = 
\frac{S_{xxxzzz}^{(6)}(r_x)}
{(S_{xx}^{(2)}(r_x))^3}
\label{eq:kurtmixed}
\end{equation}
where the numerator is a {\it
purely anisotropic} $n$th order quantity while the denominator is the 2d
order longitudinal structure function, raised to the $n/2$
power. Clearly all quantities in (\ref{eq:kurtnew}) and
(\ref{eq:kurtmixed}) 
 would be vanishing in a perfect isotropic
ensemble. The difference between the two definitions 
(\ref{eq:kurtnew}) and (\ref{eq:kurtmixed}) for $F$ and $G$, 
 lies in the normalizing
function in the denominator.
In the first case, $G$,  the normalization is through 
 a correlation of the same order of the
numerator while in the second case, $F$,  is via a  second order correlation
raised to the appropriate power. Their  amplitude as a function of $r_x$ can 
be taken as 
 a measure of the change in the  anisotropic content  as a function of 
scale.  The defintion (\ref{eq:kurtmixed}), on the other hand, 
mixes correlation of different orders,
thus including their possible different intermittent corrections 
\citep{biferale01a}. 
 In Fig.~\ref{fig5} we also show the behavior of
$F^{(n)}_{xz}(r_x)$ 
 for $n=4,6$. Again, there is a  clear indication of the presence of 
important anisotropic contributions,   particularly at small scales. 
\subsection{Probability density functions}
Before concluding this section we want to re-discuss some of the
previous results from the point of view of  the 
probability density functions (PDFs). Anisotropies may be highlighted 
at the level of the PDF by 
looking  at the antisymmetric
part of the distribution of field increments at different scales. 
Let us define the PDF,
$P(X_{\alpha\beta})$, 
of the dimensionless magnetic
field increments at scale $r_x$:
\begin{equation}
X_{\alpha\beta}(r_x) =  \frac{\delta_{r_x}B_{\alpha}  \delta_{r_x}B_{\beta}}
{\langle \delta_{r_x}B_{x}  \delta_{r_x}B_{x} \rangle}. 
\label{eq:pdf}
\end{equation}
In order to make the stochastic variable dimensionless we have
normalized it with the longitudinal second order structure functions
at that scale.  With a suitable choice of 
the indices $\alpha\beta$, all odd moments of 
$X_{\alpha\beta}(r_x)$ would be zero in a perfectly isotropic
ensemble.  This is the case when
$\alpha=x$ and $\beta=y,z$.  
We may now  define the antisymmetric part of $P(X_{\alpha\beta})$ as 
\begin{equation}
\label{eq:asimm}
A_r(X_{\alpha\beta})={P(X_{\alpha\beta}(r_x))-P(-X_{\alpha\beta}(r_x))},
\end{equation}
and notice that it would vanish in a symmetric isotropic ensemble.

\begin{figure}
\plotone{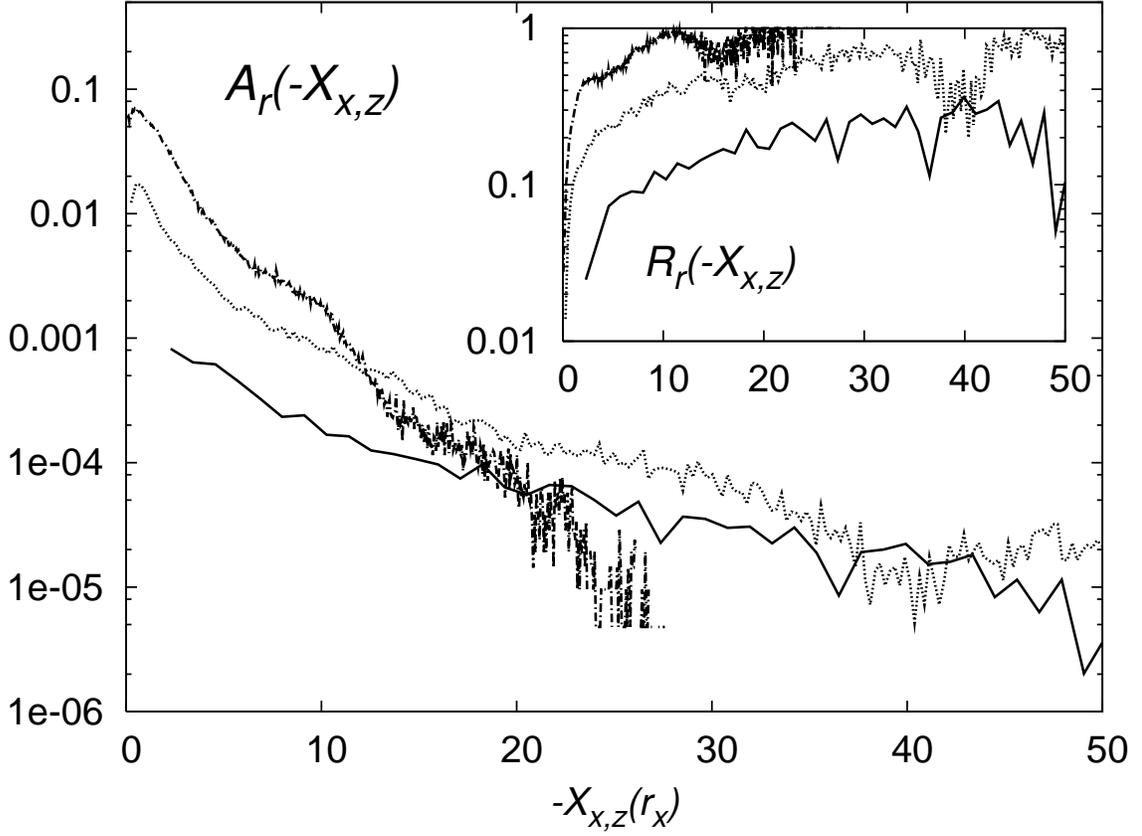}
\caption{
Antisymmetric part of the PDF of $X_{xz}(r_x)$, $A_r(-X_{xz})$ 
Eq.~(\ref{eq:asimm}), for three different spatial separations $r$.  
Solid line: $r=12$~Mm, 
dotted line: $r=192$~Mm, 
dot-dashed line: $r=3072$~Mm. 
Inset: the normalised antisymmetric part of the PDF, $R_x(-X_{xz})$,
Eq.~(\ref{eq:asimmnorm}),  for the same set of $r_x$. 
}
\label{fig6}
\end{figure}

$A_r(X_{\alpha\beta})$ gives us a
direct measurement of the  anisotropy as the 
imbalance in the probability of having oppositely
directed fluctuations at that scale.  In Fig.~\ref{fig6} we show the
antisymmetric part of the PDF, 
$A_r(X_{\alpha\beta}(r_x))$ for $\alpha = x
$ and $\beta = z$ for three different separations $r_x$. 
The increasingly fat tails as one goes to smaller scales, reflects
the non-gaussianity  of ${P}(X_{xz}(r_x))$, which becomes more enhanced at
small scales.
In order to assess the relative weight of the antisymmetric versus the
symmetric fluctuations, we define the normalised  antisymmetric part of
$P(X_{\alpha\beta})$:
\begin{equation}
\label{eq:asimmnorm}
R_x(X_{\alpha\beta})=\frac{P(X_{\alpha\beta}(r_x))
      -P(-X_{\alpha\beta}(r_x))}
{P(X_{\alpha\beta})+P(-X_{\alpha\beta})}, 
\end{equation}
This quantity also vanish in a symmetric isotropic ensemble,  
 approaching
the value one in the limit case of strong anisotropy, 
$P(X_{\alpha\beta}) \gg P(-X_{\alpha\beta})$. 
In the inset of the same figure, $R(X_{xz})$ is shown. 
The fact that at large separations $R(X_{xz})$ is close to one, means
that 
large events are progressively more anisotropic as they grow in intensity, a
possible signature of the large scale structures in the plasma. 
For small separations, the system is indeed globally more   isotropic, 
although small scale anisotropy  never vanish and survives at a significant 
level of 10\% for all intensities. 
\subsection{High-latitudes data}
\label{s:pol}
We discuss here anisotropy and intermittency detected
in the polar region by  Ulysses. This allows us
to address the ``universality'' of anisotropy, i.e. quantifying to which extent
intensities of anisotrpic fluctuations and their scaling properties
are dependent/independent on the mean large scale structure on the
magnetized plasma. 
There are two effects which might influence the relative anisotropy of
the turbulence in the polar and equatorial regions. In the polar
regions, the amplitude of turbulence relative to the mean field is
stronger, while the effects of solar rotation, which tend to bend the
interplanetary magnetic field into a spiral, are negligible. In the
equatorial high speed streams, the average magnetic field is bent into
the Parker (spiral) direction, so that there are two main axes which
may influence the evolution of the fluctuations, the radial and the
mean field directions.
The mean field direction coincides  with the radial direction for 
 polar
flows while it is perpendicular  to it, close to the $y$ direction, for the
low-latitude data arlund $5$~A.U.

\begin{figure}
\plotone{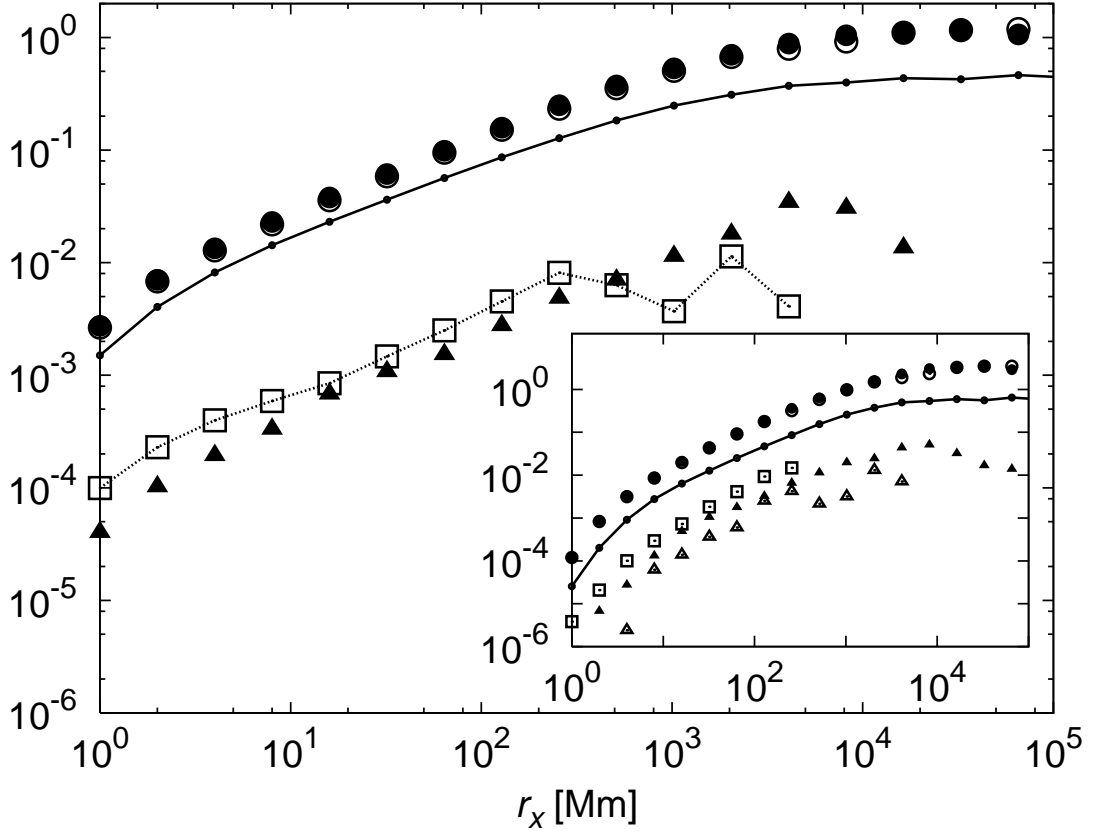}
\caption{Same as in Fig.~(\ref{fig3}). Polar dataset. 
Second order longitudinal, transverse and {\it purely
anisotropic}
structure functions. 
The upper three curves show the longitudinal and tranverse structure
functions:
solid line --- $S^{(2)}_{xx}$,
empty cirles $\circ$  $S^{(2)}_{yy}$ and filled circles 
$\bullet$ $S^{(2)}_{zz}$.
The lower curve with triangles 
 show the largest purely anisotropic structure functions: 
$S^{(2)}_{xy}$, $\blacktriangle$ filled triangles;
$S^{(2)}_{yz}$, $\square$ empty squares.
Inset: fourth order structure functions, longitudinal, transverse and {\it
purely anisotropic}.
Solid line, ---  $S^{(4)}_{xxxx}$, empty circles $\circ$ 
$S^{(4)}_{yyyy}$, filled circles $\bullet$ 
$S^{(4)}_{zzzz}$. 
{\it Purely anisotropic} structure functions are:
$S^{(4)}_{xyyy}$, $\blacktriangle$ filled triangles;
$S^{(4)}_{xzzz}$, $\vartriangle$ empty triangles;
$S^{(4)}_{yzzz}$, $\square$ empty squares.
 \label{fig7}}
\end{figure}

Let us first present results on the overall relative
importance of anisotropic fluctuation with respect to the undecomposed
ones. In Fig.~(\ref{fig7}) we show the same as in  Fig.~(\ref{fig3}) but
for polar data. {\it Purely anisotropic}
structure functions have a much lower intensity (one order of
magnitude less) with respect to the longitudinal and transverse
structure functions both for the second order (body of the figure) and
for the fourth order (inset). Indeed, for higher order moments, 6 and
higher, the statistical fluctuations combined with the very low intensity
of the anisotropic signal do not allow to have stable results even
with the whole statistic of 21 consecutive days we analyzed. We
conclude therefore that the anisotropy content at this latitude is
much lower than in the low latitude dataset. 
One could argue that at this latitudes averaging over long
periods may hide important physical phenomena which appear on a
shorter time window. Therefore, we also selected 
periods of 2-3 consecutive days when the anisotropic signal looked
more stable and intense. The anisotropic content in those events is 
 sligthly more
important and allow to make a quantitative estimate of its scaling
properties, but do not differ qualitatively.  
In Table~\ref{tab:univ}, data with an asterisk $*$ indicate that scaling
exponents are evaluated on the smaller dataset. 

\begin{figure}
\plotone{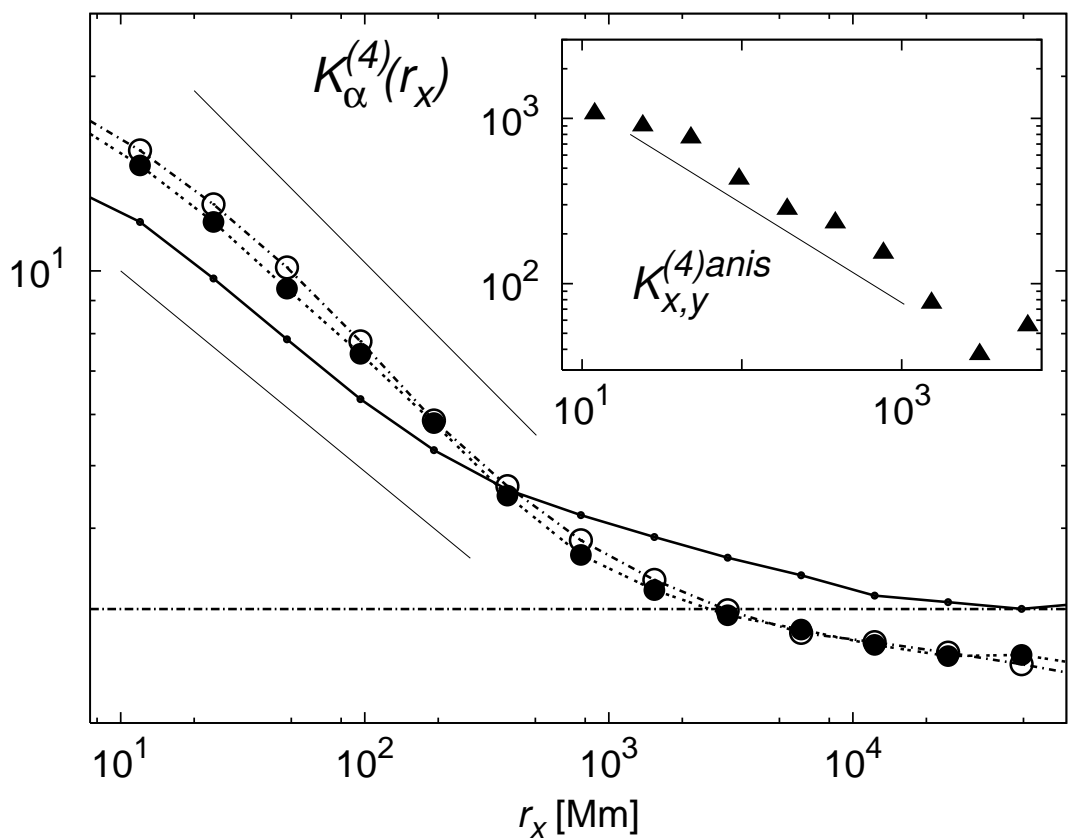}
\caption{Same as in Fig.~(\ref{fig4}). Polar dataset.
Kurtosis (\ref{eq:kurt_ls}) of the longituinal and transverse magnetic field
fluctuations.
Solid line ---  $K^{(4)}_{x}(r_x)$, empty circles 
$\circ$  $K^{(4)}_{y}(r_x)$ and
filled circles $\bullet$ $K^{(4)}_{z}(r_x)$.
The dot-dashed line shows the constant level of three for the kurtosis of a
Gaussian variable. 
In the inset: {\it purely anisotropic} kurtosis (\ref{eq:kurtanis}) of
component $\blacktriangle$ $K^{(4)}_{xy}(r_x)$.
\label{fig8}}
\end{figure}

In Fig.~(\ref{fig8}) we show the same as
Fig.~(\ref{fig4}), for the polar data set. We show the
Kurtosis of longitudinal and transverse magnetic field fluctuations
toghether with the Kurtosis for {\it purely anisotropic } correlation
functions (\ref{eq:kurtanis}). Comparing the scaling behaviours of all the
statistical indicators considered, summarized in Table~\ref{tab:univ},
we have a qualitative agreement between the polar and
the ``Equatorial'' data set. 
 If confirmed by other measurements, and/or with higher statistical data sets, this would be a nice 
indication of ``universality'' in the small scales fluctuations of the solar
wind plasma. Overall intensities of isotropic and anisotropic contents are of
course dependent on the distance and latitude, while their variation with 
scale/frequency look more stable. 
%

%
%
\begin{deluxetable}{cccccc}
\tablewidth{0pt}
\tablecaption{Universality. Scaling exponents, 
Eqs.~(\ref{eq:kurtanis},\ref{eq:zeta2},\ref{eq:zeta4})
\label{tab:univ}
}
\tablehead{
 &  $\zeta_2^{aniso}$
 &  $\zeta_4^{aniso}$
 &  $\zeta_6^{aniso}$
 &  $\chi_4^{aniso}$ 
 &  $\chi_6^{aniso}$ 
}
\startdata
Low-lat
& $ 0.75 \pm 0.15 (xy)$ 
& $ 0.8 \pm 0.3 $ ($xxxy$\tablenotemark{\dagger}) 
&  $ 1.2 \pm 0.4 $ ($xxxyyy$) 
& $ -0.6 \pm 0.2 $ ($xy$) 
& $ -1.2 \pm 0.3 $ ($xy$) 
\\ 
  &  $0.95 \pm 0.10$ ($xz$) 
  &  $ 1.0 \pm 0.15 $ ($xzzz$) 
  &  $ 1.2 \pm 0.2 (xxxzzz) $
  &  $ -0.8 \pm 0.2 $ ($xz$) 
  &  $-1.5 \pm 0.3$ ($xz$) 
\\
     &   $0.75 \pm 0.10$ ($yz$) 
     & $ 1.0 \pm 0.25 $ ($yzzz$) 
     & $ 2 \pm 1$ ($yyyzzz$) 
     &  $ -0.45 \pm 0.2 $ ($yz$) 
     &  --- 
\\
\hline
Hi-lat 
& $0.75 \pm 0.15$ ($xy$) 
& $ 0.8 \pm 0.2 $ ($xxxy^{\dagger}$)
& $ 1.1 \pm 0.3 $ ($xxxyyy$)  
& $ -0.6 \pm 0.2 $ ($yx^{\dagger}$) 
&$-1.1 \pm 0.3$ ($xy$)
\\ 
   &  ---
   & $ 0.8 \pm 0.3 $ ($xzzz$)
   & $ 1.1 \pm 0.3 $ ($xxxzzz$\tablenotemark{*})  
   &  ---
   &  ---
\\ 
      & $0.75 \pm 0.15$ ($yz$)  
      & $ 1.0 \pm 0.3 $ ($yzzz$)
      & $ 1.5 \pm 0.4 $ ($yyyzzz$\tablenotemark{*})  
      & --- 
      & ---
\enddata
\tablenotetext{*}{
Exponents evaluated on the short polar dataset, see Sect.~\ref{s:pol}.
}
\tablenotetext{\dagger}{
First component provides a third order
correlation rather than the second one, as in all other cases considered.}          

\end{deluxetable}
\section{Conclusions}
\label{s:concl}
Our main finding is
the detection of strong anisotropic fluctuations in the equatorial
part of the orbit. Here, the anisotropic contents of fourth order
correlation function is roughly of the same order of its isotropic
part, at all scales,  indicating that small scale isotropy is not
recovered.  Moreover, a high degree of
intermittency is measured in the purely  anisotropic fluctuations.  In
the polar region, anisotropies are smaller and highly fluctuating in
time, but with a spatial dependencies compatible, within statistical
errors, with the one observed at low latitudes.
This would indicate some
universal features of anisotropic solar fluctuations independently of
the latitude, at least for what  concerns their scaling properties.
Our results point toward a crucial role played by anisotropic
fluctuations in the small scales statistics. 
Models where higher order statistic is also taken into account,
providing estimates for the scaling exponents of higher order
anisotropic structure functions, will be important to a deeper
understanding of solar wind turbulence.  

Before concluding, let us go
back to the issue of distinguishing different anisotropic
fluctuations. As one learns from the theory of group of rotation in
three dimensions, there is not a unique ``anisotropic'' sector,
rather different anisotropic properties are described by projection on
the eigenfunctions with different total angular momentum, $j$, and
projections of the total angular momentum on a given axis,
$m$, \cite{arad99}.  As mentioned in the introduction, 
the exact decomposition in different
anisotropic sectors is possbile only if using numerical data, providing
access to the whole magnetic field in the 3D space. Here we have described the
procedure that should be adopted   for  one-dimensional strings of data, 
 extracting the ``whole'' anisotropic components out of the experimental data. 
This implies that all the estimate of scaling properties as here reported
 may well be affected by out-of-control 
contributions from different anisotropic sectors. The hope is that out 
of all anisotropic sectors, only the leading one is dominating the 
statistics at small scales. This hypothesis, which implies a hierachy between 
the scaling exponents in different sector   has been verified on direct
numerical simulations of
 turbulent flows \cite{biferale01,bif02} and on analytical calculation for passive magnetic fields \cite{lanotte,ref:00ABP}, but remains 
an open question for active magnetic fields.
\acknowledgments
 We are thankful to  B. Bavassano,  R. Bruno,  A. Lanotte and F. Toschi for 
 friutful discussions. 
We acknowledge support from EU under the grant ``Nonideal Turbulence'' HPRN-CT-2000-0162.


\end{document}